\documentclass{article}
\usepackage[utf8]{inputenc}
\usepackage{graphicx,caption}
\usepackage{subfig}
\usepackage{hyperref}
\usepackage{braket}
\usepackage{pdfpages}
\usepackage{color,soul}
\title{Topological defects of integer charge in cell monolayers}
\author{Kirsten D. Endresen, MinSu Kim, and Francesca Serra$^*$\\
Department of Physics and Astronomy, Johns Hopkins University\\
Baltimore, MD 21218\\
$^*$ {\normalsize francesca.serra@jhu.edu}}
\date{}

\begin{document}

\maketitle

\begin{abstract}
Many cell types spontaneously order like nematic liquid crystals, and, as such, they form topological defects. While defects with topological charge $\pm$1/2 are common in cell monolayers, the defects with charge $\pm$1, relevant in the formation of protrusions and for cell migration in living systems, are more elusive. We use topographical patterns to impose topological charge of $\pm$1 in controlled locations in cell monolayers. We compare the behavior of 3T6 fibroblasts and EpH-4 epithelial cells on such patterns, characterizing the degree of alignment, the cell density near the defects, and their behavior at the defect core. The patterned substrates provide the right conditions to observe isotropic packing of 3T6 cells in the +1 defects. We also observe density variation in the 3T6 monolayers near both types of defects over the same length-scale. These results indicate a defect core size of 150$\mu$m-200$\mu$m, with isotropic packing  possible in the +1 defects and, in other cases, a defect core with rich internal structure. 
\end{abstract}

\section{\label{sec:level1}Introduction}

The importance of nematic ordering in the alignment of living cells is gathering more evidence by the day. The idea that cell layers exhibit liquid crystal (LC) order dates back to the pioneering studies of Yves Bouligand on chitin shells \cite{bouligand1984, bouligand2008rev}, but in recent years the evidence of the connections between biology and liquid crystals has stimulated a resurgence of interest in the LC nature of cells. 

Cells confined onto 2-dimensional (2D) substrates arrange as nematic LCs \cite{ladouxreview} characterized by a strong tendency of cells to align with their neighbors. This is evident in the case of bacterial cells \cite{volfson2008} or of elongated cells such as fibroblasts or myoblasts \cite{li2017, martella2019, duclos2014sm, duclos2017}. In cases of fairly isotropic cells such as certain types of epithelial cells (or, at any rate, cells whose aspect ratio is well below the Onsager ratio), their dynamics can be treated as an active nematic system \cite{blanch2018, yeomansiso}.

As LCs, cell layers show topological defects. These are regions where the nematic order is lost \cite{chandrasekhar, kleman2006} in order to minimize stresses in the ordered fluid.  In 2D, defects are characterized by a topological charge, i.e. the angle by which LC mesogens rotate around the defect, divided by 2$\pi$ \cite{kleman2006}. This quantity is additive, conserved, and determined by the topology of the LC confinement. For example, if nematic LCs are confined on the surface of a sphere, the total topological charge is +2 and it can be obtained with an arbitrary number of defects whose charges add up to +2. In nematics, topological charge can be integer or semi integer, the most common defects being $\pm$ 1/2 and $\pm$ 1. In low molecular weight LCs, defects interact with each other strongly by elastic interactions, they are able to trap colloidal particles and small molecules \cite{lacaze, abbott, senyuk}, they exhibit interesting optical effects \cite {me, brasselet} and in general they are mediators of self-assembly \cite{musevic}. 

Cell layers also form topological defects as they rearrange, and it is becoming evident that these defects have a biological role. Just as in active LCs \cite{dogic, giomi2013, giomi}, the defects with topological charge +1/2 drive the dynamics of the cell layer, having a strong elastic dipole, while the defects with charge -1/2 are moved around passively  \cite{yeomans2014}. Saw et al. \cite{saw2017} found that near +1/2 defects the rate of apoptosis of MDCK epithelial cells is higher, due to the presence of isotropic compressive stresses. In contrast, the -1/2 defects are characterized by tensile stresses and do not trigger apoptosis. Kawaguchi et al. \cite{sano} report a strong effect of topological defects in murine neural progenitor cells moving on substrates without attaching. The density of the monolayer increases and eventually stops moving next to the +1/2 defects, resulting in the formation of large cell conglomerates. Similar conglomerates have been reported in bacteral systems \cite{yeomans2016bacteria}. The importance of +1/2 defects seems consistent across a broad spectrum of cells, and it could be responsible for the formation of epithelial extrusions or other 3D structures \cite{ladouxreview, ladoux2015, trepat2018mesoscale}. 

Defects with topological charge $\pm$ 1/2 are commonly observed in cell culture and have therefore been the subject of many studies. On the other hand, defects with integer charge are not typically observed in flat layers and have received less attention. However, these types of defects are still present in living systems. Examples of topological defects with +1 charge (Figure 1a) arising in tissues include cells around the optic nerve \cite{gogola} and at the tumor-stromal interface \cite{provenzano2006collagen}. Bade et al. showed how the cytoskeleton rearranges when cells are near a protrusion \cite{nathan}, leading to dynamics consistent with either radial or azimuthal +1 defect. The collagen fibers at the tumor-stromal interface cause cells to align radially and promote migration \cite{provenzano2006collagen}. A circular alignment has also been seen in the formation of rosettas of epithelial cells during extrusion \cite{rosenblatt}. 

Here, we use topography to investigate the behavior of cells near induced defects with topological charge +1 and -1, to (a) characterize how well cells can adapt to these undesirable distortions, (b) characterize the alignment of cells close to the defects and (c) extract important physical LC parameters of the in-plane ordering of cell monolayers. In order to achieve this, we use topographical patterns with micron-high ridges to influence the local alignment of cells by gentle cues without entirely confining them. 

In this paper, we focus on fibroblasts and epithelial cells, chosen as epitome of two different cell types: fibroblasts interact strongly with the substrate and are able to assume very anisotropic shapes; in contrast, the EpH-4 epithelial cells have strong cell-cell junctions and are quite isotropic in shape. Our patterns allow us to investigate these situations where cells monolayers experience the frustration of undesired topological structures, thereby gathering information on the LC behavior of the cell layers.

\section{\label{sec:level1}Results}
We analyze the behavior of cells on a PDMS pattern that creates a square array of +1 and -1 defects. We vary the spacing between the +1 and -1 defects and also the distances between the ridges of the pattern. The width of the ridges is always 9$\mu$m with a height of 1.5 $\mu$m. The details of the fabrication are in the Experimental section. We observe 3T6 fibroblasts from albino mice on patterns where the spacing between ridges, $r$, is 60, 90 and 120 $\mu$m. The results are shown in Figure 1, where we measure the angle of alignment of the cells (Fig. 1b). The angles are measured from the center of the topological defects. As expected, the fibroblasts follow the ridges (Fig. 1c-g), and they can align well on all three patterns. They show the expected orientation around the defect as shown in Figure 1e-f, where every point of the diagram represents the orientation of a cell with respect to their nearest +1 or -1 defect, and the slope depends on the charge of the topological defect. The cell orientation is obtained by staining the cells with cell-permeant nuclear stains (Hoechst 33342 and NucRed Live 647), fitting their nuclei shape to ellipses, and analyzing their orientation. Details are included in the experimental section. Figure 1g shows a similar angle distribution of the cells for every pattern. It is interesting to notice that in the case of the +1 defects there is a population of cells oriented perpendicular to the expected direction. In fact, we observe that some cells form bridge structures across the ridges and orient perpendicularly.

\newpage

\begin{figure}[h!]
\begin{center}
	\includegraphics[width=0.7\textwidth]{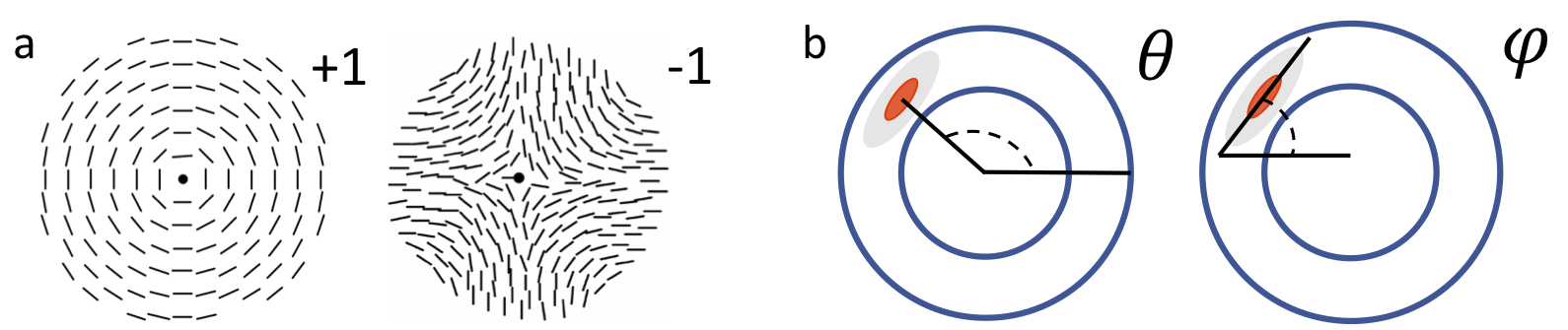}
	\end{center}
    \includegraphics[width=0.9\textwidth]{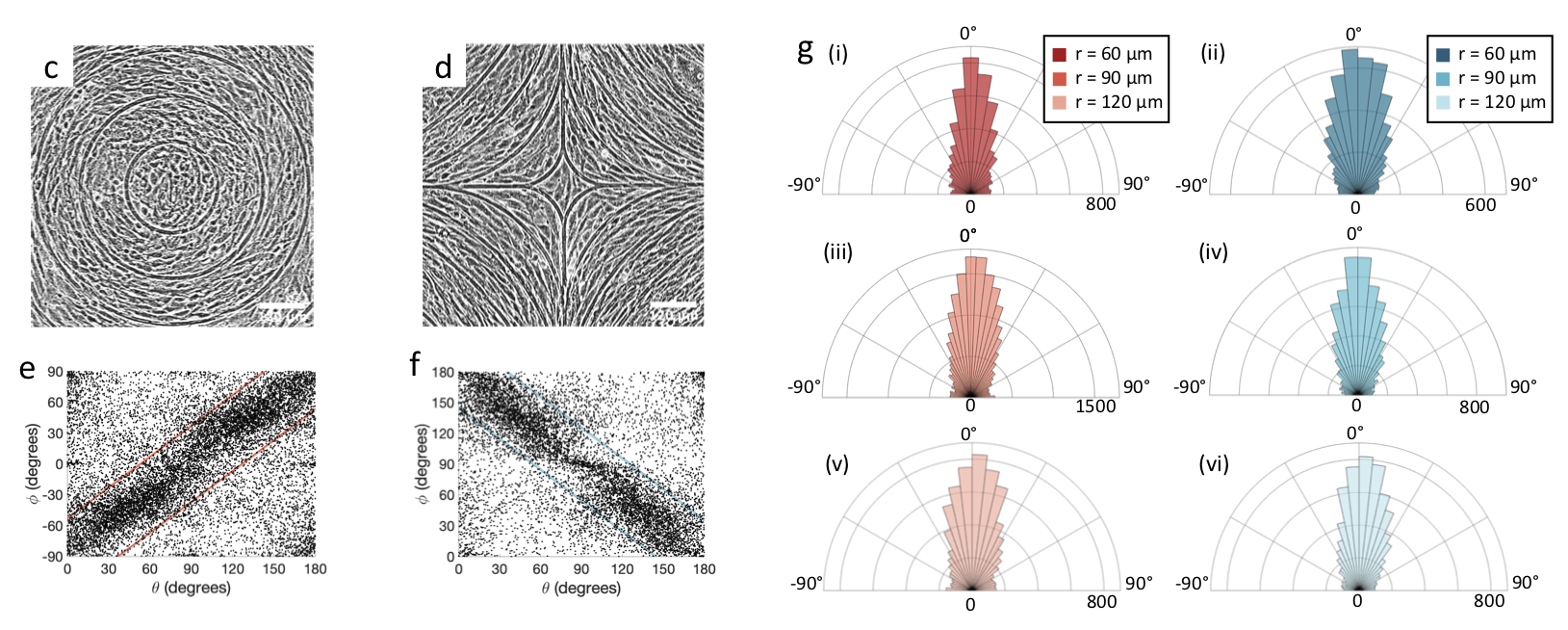}
    \caption{(a) Schematic of a +1 azimuthal and a -1 topological defect. (b) Schematic of $\theta$, corresponding to the angle from the center of the defect, and $\phi$, the angle of the major axis of the nucleus.(c) Phase contrast (PC) image of 3T6 cells in the vicinity of a positive defect on r=120$\mu$m pattern. Scale Bar is 120$\mu$m. (d) PC of 3T6 in the vicinity of a negative defect on r=120$\mu$m pattern. Scale bar is 120$\mu$m. (e) Scatter plot of 3T6 alignment with r=90$\mu$m positive defect pattern. Red lines indicate root-mean-square (rms) deviation from the expected angle, with the shade corresponding to (g), indicating the rms value for all the patterns. In this case, they overlap. (f) Scatter plot of 3T6 alignment with r=90$\mu$m negative defect pattern. Blue lines indicate rms deviation in angle for all three patterns, corresponding to the shades in (g). Alignment of 3T6 with ridges for (i-ii) r=60$\mu$m, (iii-iv) r=90$\mu$m, and (v-vi) r=120$\mu$m patterns around (i, iii, v) positive defects and (ii, iv, vi) negative defects. Radial axis represents number of cells, and angle represents the deviation of cell alignment from the expected value. 
\label{alignment}}
\end{figure}

\begin{figure}[h!]
    \includegraphics[width=0.9\textwidth]{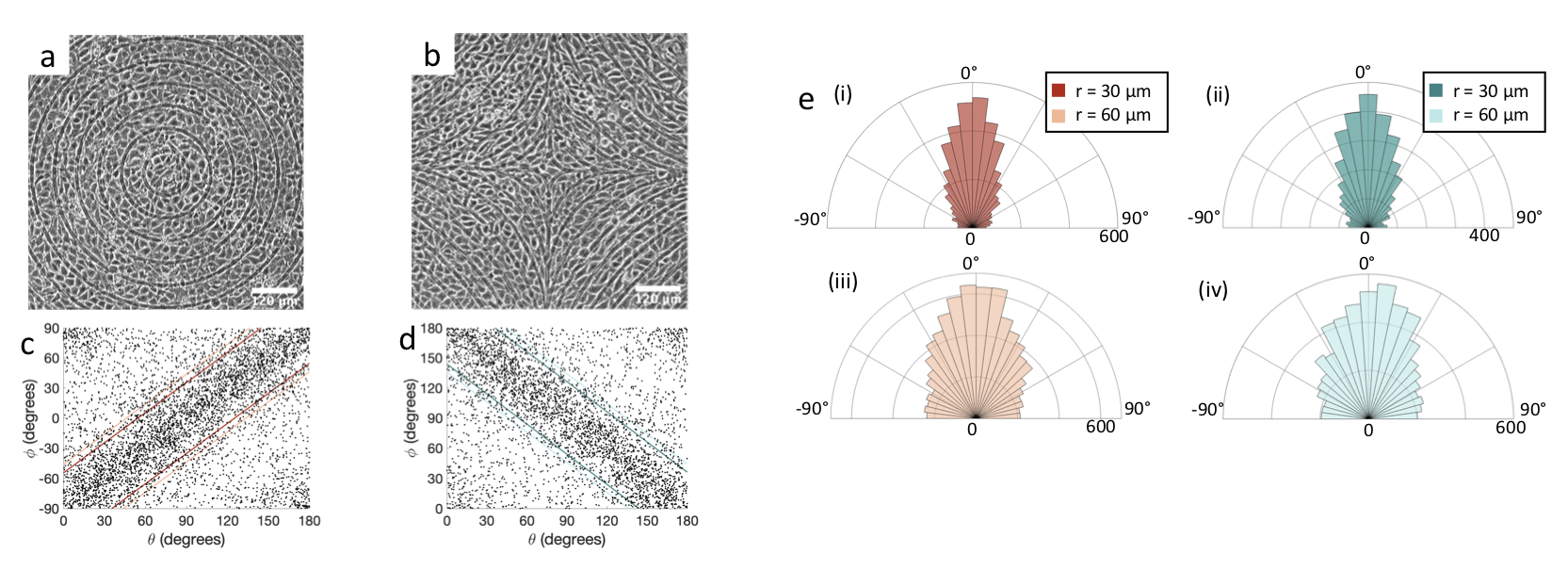}
\caption{Alignment of EpH-4 cells near defects. (a) PC of EpH-4 cells in the vicinity of a positive defect on r=60$\mu$m pattern. Scale bar is 120$\mu$m. (b) PC of EpH4- cells in the vicinity of a negative defect on r=60$\mu$m pattern. (c) Scatter plot of EpH-4 alignment with r=30$\mu$m positive defect pattern. Orange lines indicate rms values for r=30$\mu$m and r=60$\mu$m. (d) Scatter plot of EpH-4 alignment with r=30$\mu$m negative defect pattern. Green lines indicate rms values for r=30$\mu$m and r=60$\mu$m. (e) Alignment of EpH-4 with ridges of (i-ii) r=30$\mu$m and (iii-iv) r=60$\mu$m patterns around (i,iii) positive defects and (ii,iv) negative defects.
\label{alignment2}}
\end{figure}

We perform the same experiment with epithelial cells (Fig. 2). If we use the same patterns used for the fibroblasts the alignment observed in the cells is present, but much weaker. In Figure 2 we show only the results for the 60 $\mu$m pattern (Fig. 2a-d). We then utilize a pattern with ridges which are more finely spaced, with 30 $\mu$m spacing, and indeed that gives a much better alignment to the epithelial cells, comparable to that of fibroblasts, as can be seen from the angle distribution in Fig. 2e. We notice, however, that the tighter confinement means that only 1-2 cells can fit in between adjacent ridges. 

We investigate the effect of confluency on the alignment on fibroblasts. The fibroblasts achieve the best alignment with cells when they initially become dense enough to be a monolayer (Supplementary Figure S1a-b). As the cell density continues to increase, the fibroblasts begin to squeeze, becoming smaller in area and more disordered. Even so, the epithelial cells, with an even smaller aspect ratio, do not achieve a similar degree of alignment. 

Having confirmed that the cells can align on our patterns, we track the cell density $\rho$, defined as the number of cells per area, in the vicinity of the topological defects. We then consider the relative densities of the cells at various distances from the center of the defects, each normalized by the average sample density. In the case of fibroblasts, every pattern indicates that the density is maximum near the +1 defect and minimum near the -1 defects (Figure 3a-b). This is consistent with observations made on cells near +1/2 defects \cite{sano}. The difference in each case is more marked for the samples with the smallest spacing between ridges. This effect is very evident in the +1 defects when the cells are tightly packed. We observe, in fact, that for lower density of fibroblasts (1000cells/mm$^2$, VS the typical c.ca 1500cells/mm$^2$ for the patterns in Figure 3a-b) the density variation near the +1 defect is much less marked (Supplementary Figure S1c), while the density decrease near the -1 defect is still present (Supplementary Figure S1d). The case of epithelial cells is different. Epithelial cells show a slight increase of density near both defects of positive and negative charge (Figure 3c-d), and also in this case the difference is more marked for the finely spaced pattern. Consistently, the work of Saw et al. \cite{saw2017} showed an increase of apoptosis near +1/2 defects not correlated with an increase in cell density of epithelial cells.   


\begin{figure}[h!]
    \includegraphics[width=0.6\textwidth]{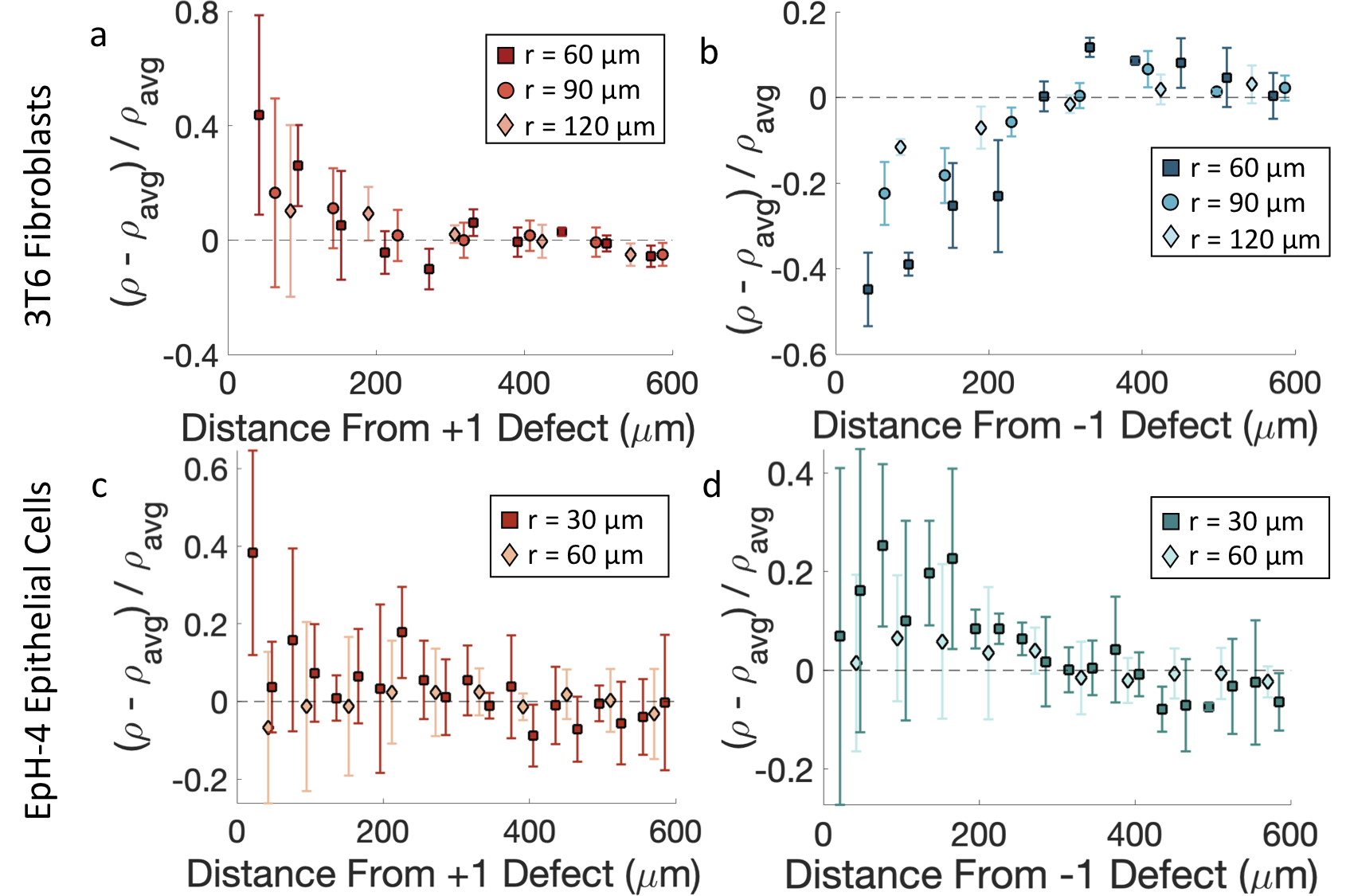}
\caption{Density of cells near defects with positive and negative topological charge. (a-b) Density of fibroblasts near +1 (a) and -1 (b) defects. The data refer to patterns with three different sizes. (c-d) Density of epithelial cells near +1 (c) and -1 (d) defects. All densities are normalized by the average density of each pattern.}
\end{figure}

The results shown in Fig. 3a-b lead to two conclusions: that the behavior of fibroblasts depends on the charge of the defects and that the lengthscale of the pattern is relevant. To understand this, we observe more closely what happens in the inner region near the defects. Typical results are shown in Figure 4a-c. If the inner circle has a radius above 100 $\mu$m, the cells tend to arrange in a polar configuration with two +1/2 defects near the edges of the circle. This was also shown by \cite{duclos2014sm, duclos2017} where the fibroblasts were confined in round patches with radii of several hundred microns. However, if the radius of the inner circle is reduced, the cells can adopt a different configuration: they become more round and they pack isotropically. The graph in Figure 4d shows the relative prevalence of the +1 defect with respect to the two +1/2 defects as a function of the feature size. It is immediately noticeable that the fraction of +1 defects has a maximum for radii around 60-80$\mu$m but is lower at smaller radii. This is easily explained. The cells, too tightly confined, prefer to ignore the alignment cues from the ridges and form two +1/2 defects at a larger distance, as shown in Figure 4a.

\begin{figure}[h!]
    \includegraphics[width=0.6\textwidth]{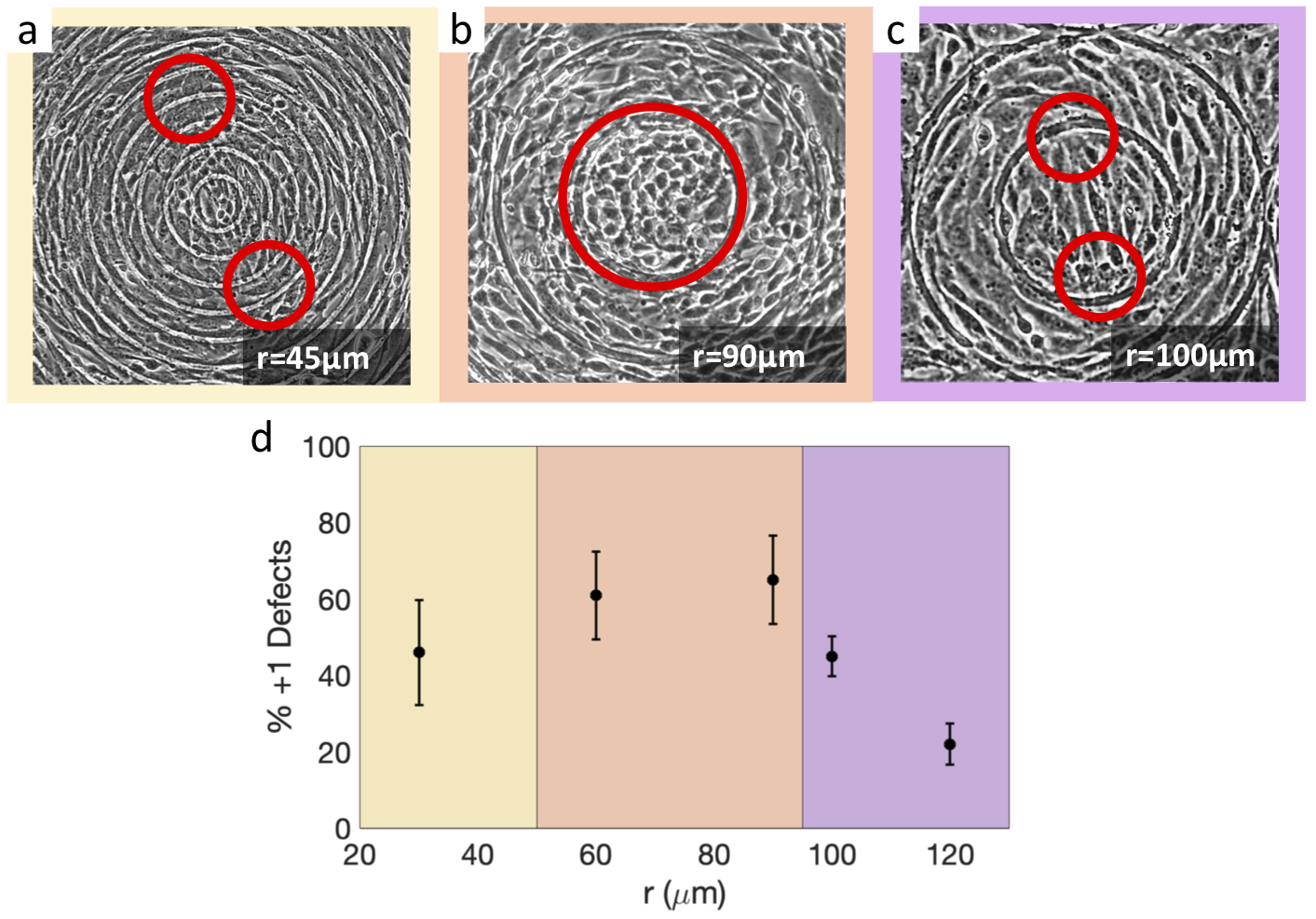}
\caption{Examples of cell alignment near the +1 defects for fibroblasts. (a) Defect splitting into two +1/2 defects far from the inner ring, (b) isotropic region in the inner ring and (c) defect splitting into two +1/2 defects in the inner ring. (d) Relative frequency of isotropic regions VS polar configurations with two +1/2 defects.}
\end{figure}

Fibroblasts near defects with charge -1 are harder to analyze because cells do not pack densely near it. Therefore we observe less cells, especially for the more finely spaced ridges. For every pattern size, however, the -1 defects split into two -1/2 defects whenever the cell density is high enough to form a monolayer (Figure 5). In all cases, the cells in the central part of the defect are oriented in either the north-south or east-west directions (Figure 5a-b). Only in very few cases we observe a structure resembling a -1 defect (Figure 5c). 

The case of epithelial cells is significantly different. The alignment of cells is not greatly perturbed in the inner regions of the defects. For +1 defects, in the 30$\mu$m patterns only few cells can fit in the inner circle, and they are always arranged isotropically. We do not see the tendency we saw in fibroblasts, i.e. the formation of two +1/2 defects near the outer ridges. As we increase the size of the inner circle to a radius of 60$\mu$m, epithelial cells are still isotropic (Supplementary Figure S2a-c). Occasionally, however, we also observe either a slight polar order in the inner circle or a rosetta-like structure, with cells arranged in roughly concentric rings. Increasing the size of the inner circle to 100$\mu$m radius, the behavior remains similar to the 60$\mu$m case. In the -1 defects the epithelial cells always maintain an isotropic arrangement (Supplementary Figure S2d-e).

\begin{figure}[h!]
    \includegraphics[width=0.6\textwidth]{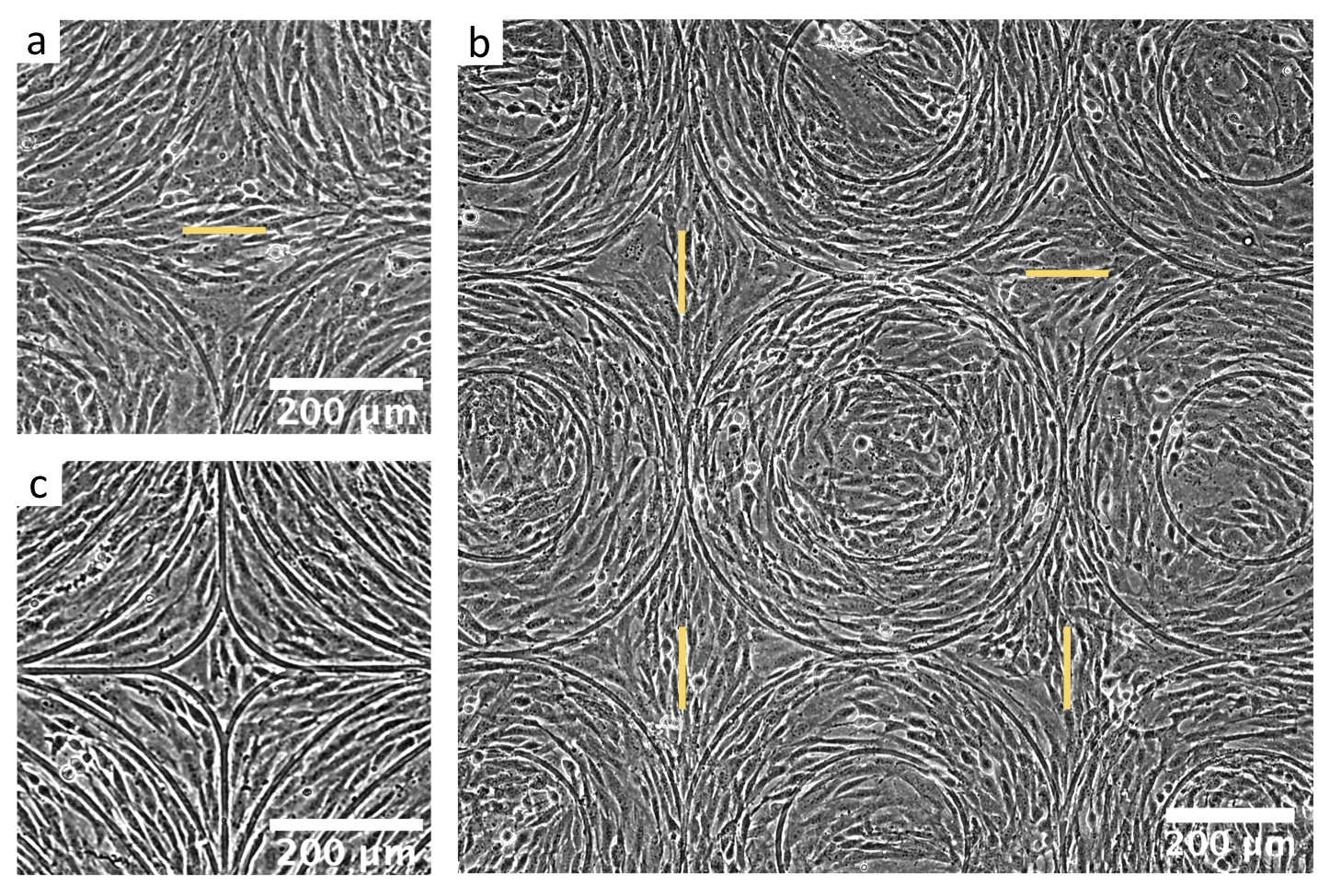}
\caption{Defects with topological charge -1.  (a) Defect clearly split into two -1/2 defects. (b) Four neighboring -1 defects, each split into two -1/2 defects either along NS or EW direction, as highlighted by the yellow rods. (c) Rare example of isotropic alignment in the inner part of the -1 defect.}
\end{figure}

We then reduce the number of topographical cues. For this, we use a pattern with only one or two ridges per circle (Supplementary Figure S3). If the circular ridge is smaller than about 40$\mu$m, the cells ignore the topographical cues and grow all over the ridges, showing random +1/2 and -1/2 defects (Supplementary Figure S3a). If the circles are 100$\mu$m or larger, the results are like those shown in Fig. 4 and Fig. 5. Also in that case, we observe the two possible configurations for the defect within the inner ring (polar or isotropic) and the splitting of the -1 defects into two defects (Supplementary Figure S3b). 
\section{Discussion}

The results presented above allow us to start a characterization of the LC properties of cell monolayers. The data presented in figure 4 and 5 give insight on the nature of the topological defects. The search for the structure of the defect core is a very hard task in thermotropic LCs, mostly studied with theoretical models and computer simulations \cite{sluckin, deluca}. The defects, however, have been observed in greater detail in other systems such as chromonics \cite{shuang} and LCs made of viral rods \cite{zazad}. In describing nematic LCs with the Landau-DeGennes theory, the free energy is described as a sum of the elastic energy, represented by the Frank energy and dependent on the elastic constants, and the phase energy which is expressed as $F_{Phase}=A(T-T^*)Q^2-BQ^3+CQ^4$, where Q is the scalar order parameter and A, B and C are three constants which depend on the material \cite{degennes}. The defect core size depends on the ratio between the elastic constant K and the term $A(T-T^*)$, as $r_C=\sqrt{\frac{8\pi K}{A(T-T^*)}}$. In typical thermotropic LCs, this size is a few tens of nanometers, corresponding to about 10 times the molecular size. In modeling LCs, the defect core is usually defined as an isotropic region within the ordered fluid. However, studies in chromonics and other lyotropic LCs, where the core size is larger, have revealed a rich internal structure \cite{shuang}. The case of the cells, in this sense, is quite special, as we see both defects with isotropic core and defects with rich internal structure. The +1 defect is the only case where we observe the isotropic arrangement of fibroblasts (Fig. 4), and this occurs for pattern size below a certain threshold, which we can identify as the core size. In all other types of defects, in contrast, we observe the internal structure. For fibroblasts, the core linear size is around 150$\mu$m, corresponding to about 10 cells, in analogy with thermotropic LCs. This characteristic length-scale is the same over which we observe a variation in the cells density (Fig. 3). We can therefore start characterizing the core size as this length-scale where the transition takes place.

As we showed in Fig. 5, the -1 defects always have an internal structure characterized by a mirror symmetry along either the north-south or the east-west direction. A crucial difference between the +1 and the -1 is that the -1 defect intrinsically has a four-fold symmetry that is also imposed in our pattern. This provides two choices of an easy axis for the cells to orient. Also reducing the distance between the -1 defects down to 400$\mu$m, we could not find any correlation between the orientation of adjacent -1 defects in our array (Figure 5b).

Although we see alignment of epithelial cells with the patterns, we do not see the same structures of the defects. This suggests that the lack of a clear transition from isotropic to polar in epithelial cells can explain their less marked density increase near the defect. We do not see an isotropic defect core of densely packed cells but we observe central symmetry of cells in the +1 defects, resembling a rosetta structure, which can either suggest that the defect core is very small (1-2 cells) or that it has a rich internal structure.

\section{Conclusions}

We have characterized the behavior of epithelial cells and fibroblasts on topographical patterns that are imposing defects of integer topological charge, not typically observed in cell monolayers. Importantly, the topographical patterns are not insurmountable barriers, but gently confining shallow ridges, which can be easily overcome by the cells. While both cell types show alignment with the ridges, the alignment is much greater in fibroblasts. Fibroblasts also show an increase of density near defects with positive charge and a decrease near defects with negative charge. The behavior of fibroblasts near the defect with charge +1 has allowed us to quantify the defect core size, which is related to the ratio between the elastic constant and the phase parameter in the Landau free energy. This characterization is made more difficult for epithelial cells due to their more isotropic nature, which creates a restriction in the confining length-scales that we can probe. 

These studies are a first characterization of the behavior of cells near defects of topological charge $\pm$1, which are hard to see in monolayers without confinement, but are observable in living systems near protrusions or along aligned fibers. There is increasing evidence not only that topological defects affect cell behavior, but also that the effects are dependent on the topological charge. We suggest that these designs for confining structures allow for the systematic study of defects with arbitrary charge and symmetry and the behavior of different cell types near them.

\section{Methods} 

{\bf Cell Lines and Culture Methods:} 
We used 3T6 fibroblasts and EpH4 epithelial cells. Cells were cultured in CellTreat Tissue Culture Dishes using 90$\%$ DMEM and 10$\%$ FBS. Cells were simultaneously growing on regular Petri dishes to verify that their growth and behavior was regular and healthy. 

{\bf PDMS Preparation:} 
We prepared polydimethylsiloxane (PDMS) substrates with 15$\%$ curing agent. Once mixed with the curing agent, PDMS was degassed with vacuum at room temperature for 20 minutes. We then poured the PDMS over the SU8 patterned substrates. These were left in a vacuum oven between 35-50$^\circ$C overnight, then for 1 hour at 80$^\circ$C to finish curing. The patterned PDMS was then prepared for cell culture. The height of the ridges was measured by cutting off a cross-section of the pattern (Supplementary Figure S4).

{\bf PDMS Preparation for Cell Culture: }
We sterilized the patterned PDMS by submerging it in ethanol for 20 minutes. Then we washed the substrates with a balanced salt solution (PBS). Fibronectin was added to the substrates as an attachment factor at 25 $\mu$g/mL, which were coated with minimal volume for 45 minutes and washed with PBS prior to use for cell culture.

{\bf Staining and Imaging Cells:}
The cells were stained using either of two cell-permeant nuclear stains: Hoechst 33258 dye (10 mg/mL stock solution purchased from Invitrogen) and NucRed Live 647 ReadyProbes Reagent (purchased from ThermoFisher). The Hoechst solution was diluted at a ratio of 1:1000 in PBS, and cells were coated with this and placed in an incubator for 15 minutes. Then the dye was removed and washed away with PDMS, and the cells were ready to image. For the NucRed stain, 20 drops were added to the 10mL of media in which the cells were growing, followed by 15 minutes in the incubator. Imaging was performed with an inverted microscope Nikon Ti-Eclipse, equipped with a Hamamatsu Orca-flash camera. At each location, two images were taken, one in phase contrast and the other with a fluorescent filter compatible with the Hoechst stain.

{\bf Quantifying Orientation of Cells:}
The phase contrast image was used to locate the topographic features. From the phase contrast image, the center of each ring was identified using ImageJ and used to define the location of a +1 topological defect. The location of a -1 defect was defined by the center point between four surrounding +1 defects. 

The orientation of nuclei was used as a proxy for the cell orientation. In order to quantify the orientation of the nuclei, we used ImageJ to first create a binary mask which showed contained only the cell nuclei, minimizing background features as much as possible. The watershed method was used to separate nuclei which appeared too close to each other to be distinguished in the image. The Fit Ellipse function in ImageJ was used to determine the center locations, major and minor axis lengths, major axis orientations, and areas of each nucleus. Ellipses were only fit to particles with areas $>$50 pixels in order to eliminate noise from the images. This method leads to under-estimating the number of cells in the very high density region as shown in Figure S5, where we present an alternative method that confirms our results.

The angle from the center of the defect was determined based on the location of the center of the defect defined from the phase contrast image and the location of the center of the nucleus given by the ellipse fitting. 

{\bf Uncertainty Analysis:}
The degree of cell alignment with the pattern was determined by comparing $\phi$ to the expected angle which, in the case of the +1 azimuthal defect pattern, was defined as $\theta$+90, while for the -1 defect pattern, the expected angle is equal to 180-$\theta$. The deviation from the expected angle we call $\alpha$, and to quantify the overall alignment of cells we compute  $\sqrt{\langle \alpha^2 \rangle}$. 

The uncertainty reported in the observed variations from the average density was reported as the range of observed values as the most conservative estimate. This was determined to be most appropriate due to the size of the data sets. For the 60 $\mu$m pattern, 3 instances of the pattern were considered for the positive defect as well as for the negative defect. For the 90 $\mu$m pattern, 7 instances of the pattern were observed for the positive defect, and 3 instances for the negative defect. For the 120 $\mu$m at high confluency, 4 instances of the pattern were considered, both for the positive and negative defects. 

The uncertainty in the defect prevalence is the uncertainty in a binomial distribution, $N\sqrt{p(1-p)}$. 

\bibliographystyle{naturemag}

\section{Acknowledgements}
The authors acknowledge Yun Chen and Wei-Hung Jung for providing the epithelial cells. This material is based upon work supported by the National Science Foundation Graduate Research Fellowship under Grant No. DGE-1746891 awarded to K.D.E. 

\section{Authors contribution}
K.D.E. performed the experiments and the analysis. MS.K. collaborated to the experiments, trained and assisted K.D.E. in the preparation of the patterns. F.S. led the project. All the authors wrote the paper.



\section*{Supplementary material}

# Topological defects of integer charge in cell monolayers

Kirsten Endresen, MinSu Kim, Francesca Serra

## Supplemental Materials

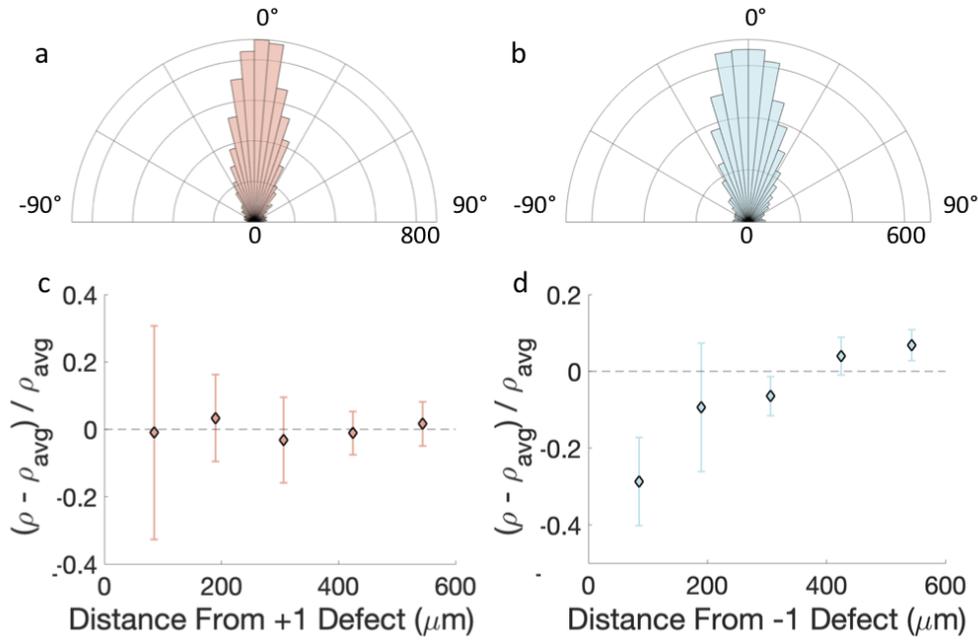

Figure S1: Alignment and density of fibroblasts on topographical patterns as soon as they reach confluency. The cell density is 1000 cells/mm$^2$ for the +1 defect and 900 cells/mm$^2$ for the −1 defect, lower compared to the value of 1900 cells/mm$^2$ used in Figure 1 (for both positive and negative defects on the r=120$\mu$m pattern). (a-b) show the angular distribution of the cells around the +1 (a) and −1 (b) defect. (c-d) show the density variation near the +1 (c) and −1 (d) defect. The rms $\alpha$ for the +1 defect under these conditions was 29°, compared to 38° for the 120$\mu$m pattern at higher confluency, detailed in Figure 1 and Figure 2. This was the best alignment achieved overall.

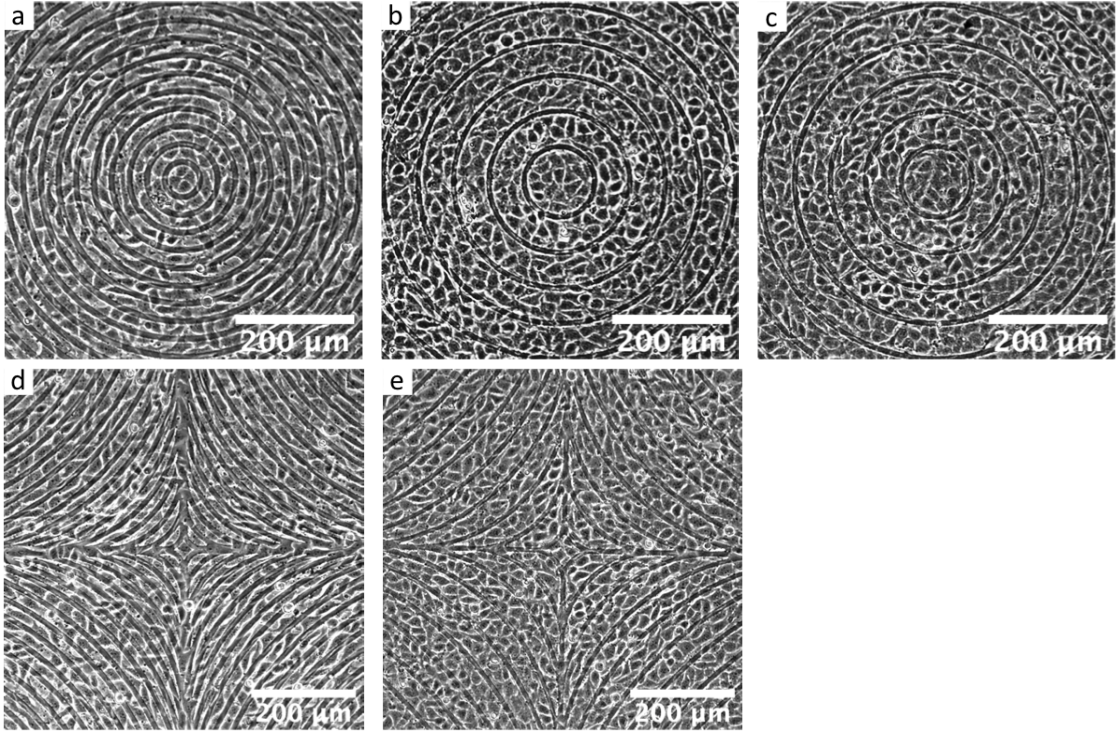

Figure S2: (a-c) Epithelial cells near the inner circle. (a) For small patterns, the cells pack tightly and isotropically. (b,c) For larger circles, they sometimes show a layered structure (b), and sometimes a polar structure (c). (d-e) Epithelial cells in the center of a -1 defect, with the small pattern yielding improved alignment (d) compared to (e).

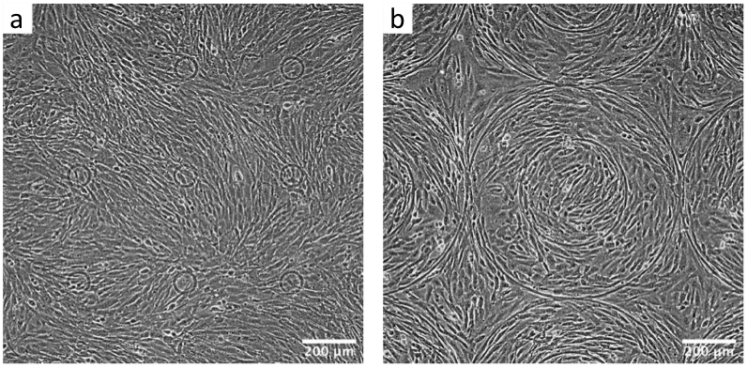

Figure S3: (a) Fibroblasts growing on small circular ridges. Fibroblasts can be seeing going over the ridges and ordering randomly throughout. (b) Cells on the $r = 200 \mu m$ two ring patterns.

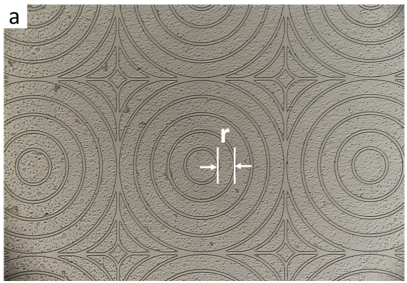 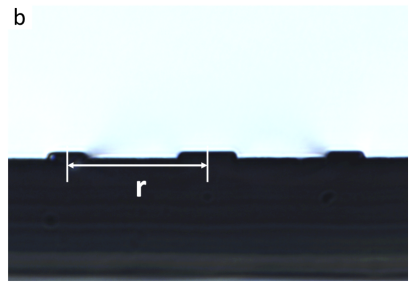

Figure S4: (a) Top view of patterned PDMS, where r is the spacing between ridges. (b) Cross section of the ridges of the PDMS pattern.

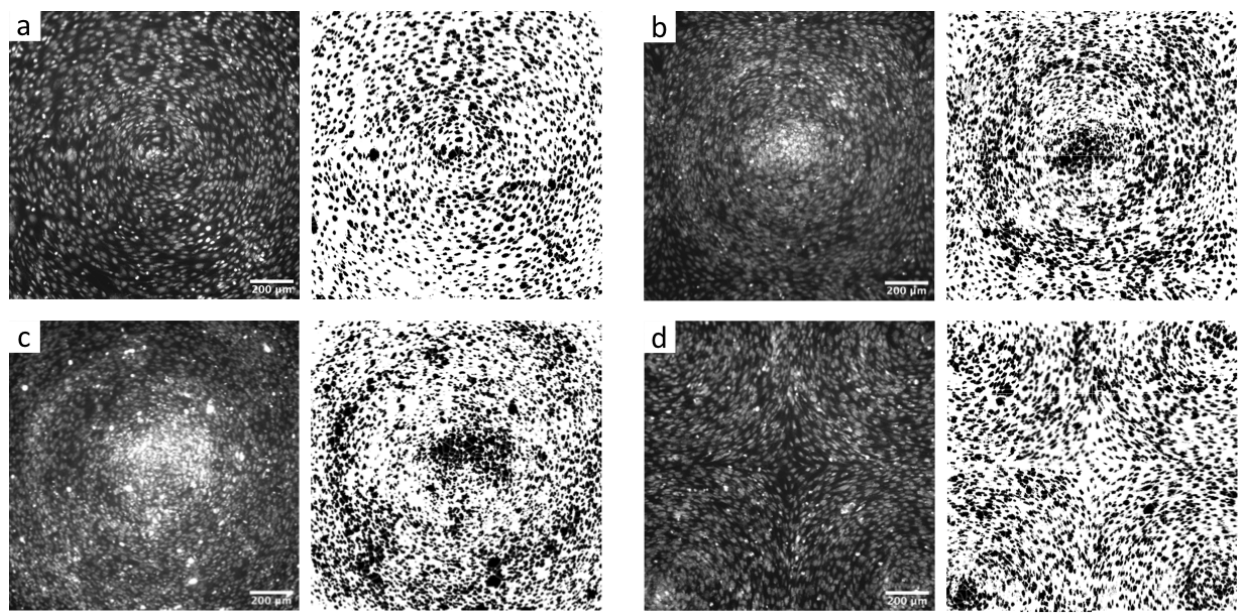

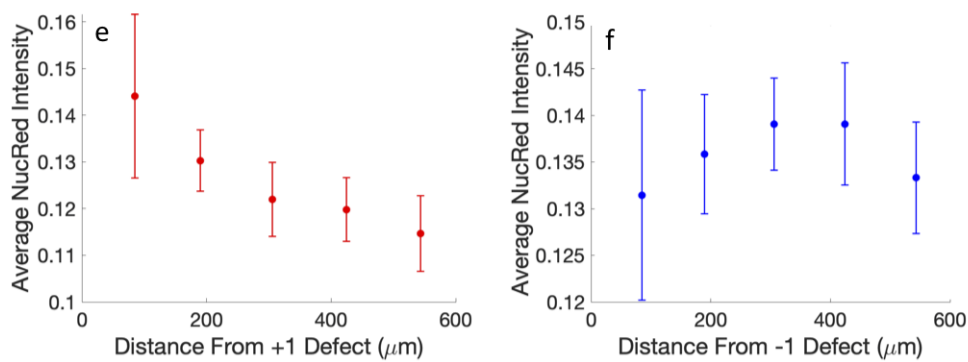

Figure S5: (a)Fluorescent image of 3T6 nuclei when cells were at lowest density considered to be confluent (1000 cells/$mm^2$) on left, with binary mask comparison on right which was used to fit ellipses on right. (b) Fluorescent image of 3T6 at higher density (1500 cells/$mm^2$) on left with binary mask comparison on right. (c) (left) Fluorescent image of 3T6 at highest density (1900 cells/$mm^2$) and (right) binary mask comparison. (d) 3T6 growing in a $-1$ defect pattern at a density of 1400 cells/$mm^2$ for comparison. The method of analysis used, represented in the right images, can fail to identify boundaries between cell nuclei at very high densities, and thus underestimates the number of cells in these regions. (e, f) show an alternative method to capture the density variation at high density with (e) showing the variation near a +1 topological defect and (f) showing the variation near a -1 topological defect. In this method, we consider the average intensity in the fluorescent images within the same regions of interest. The benefit of this method is that it is not affected in the same way at the highest cell densities, but the drawback is that it is not a direct means of obtaining the cell density. However, using this method we recover the behavior which can clearly be seen in (b, c), in which the cell density is highest and thus the image is brightest, near the center of the +1 topological defect.

14



\end{document}